\documentclass[equation,12pt]{article}
\def\lsim{<\kern-2.5ex\lower0.85ex\hbox{$\sim$}\ }
\def\rsim{>\kern-2.5ex\lower0.85ex\hbox{$\sim$}\ }

\def\ni{\noindent}

\begin{document}

\baselineskip 18pt

\centerline{\bf\Large A \lq\lq RUNNING" GRAVITATIONAL CONSTANT ?}

\vspace{.25in}

\centerline{A.C. Melissinos}

\centerline{\it Department of Physics and Astronomy, University of 
Rochester, Rochester, NY 14627}

\centerline \today \vspace{.25in}

 \ni {\bf Abstract:}

If the gravitational interaction is unified with the electro-weak
and strong interactions at a mass $M=10^{15}$ GeV, the evolution
of Newton's constant must differ from  its classical (general
relativistic) form.  We can model such behavior by introducing an
ad hoc dependence on $\ell n(s/4m^2)$, where $s$ is the usual cm
energy between two protons.  We can then predict the observable
effects for relativistic collisions  $(\sqrt s \sim 1.4 \times
10^4$ GeV) as well as for the case of low velocity motion
$(\beta^2\sim 10^{-5})$.

\vspace{1in}

It is well known that the dimensionless coupling constants of the
three gauge groups of the standard model $SU(3)\times U(2)\times
U(1)$ vary with the momentum transfer of the interaction [1]. This
effect which is due to the polarization of the vacuum was first
recognized for the electromagnetic field.  It is most prominent in
the case of the color field and leads to assymptotic freedom.

Extrapolation to higher energies is governed by the equations of
the renormalization group, and it is customary to consider the
inverse coupling constants

{\begin{eqnarray*}
 \frac{1}{\alpha_1{(\sqrt s)}}  & = &
\frac{1}{\alpha_e(m_Z)} \frac{3}{5} \cos^2\theta_W -
\frac{1}{12\pi} (4n_f)\ell
n\left(\frac{s}{m^2_Z}\right)\\
\\
 \frac{1}{\alpha_2(\sqrt s)} & = & \frac{1}{\alpha_e(m_Z)}
\sin^2\theta_W + \frac{1}{12\pi}\left(22-4n_f-\frac{1}{2}\right)
\ell n\left(\frac{s}{m^2_Z}\right)\\
\\
 \frac{1}{\alpha_3(\sqrt s)} & = & \frac{1}{\alpha_s(m_Z)} +
\frac{1}{12\pi} \left(33-3n_f\right) \ell
n\left(\frac{s}{m^2_Z}\right)
\end{eqnarray*}

The above expressions have been normalized at a cm energy $\sqrt
s=m_Z$ where the couplings are given by

\vfil\eject

\begin{eqnarray*}
& &\alpha_e   =  \frac{e^2}{(4\pi\epsilon_0)\hbar c} =
 \frac{1}{128}\\
 \\
& &\alpha_3  =  \alpha_s = \frac{g^2_s}{\hbar c} = 0.118\\
\\
& &\sin^2\theta_W(m_Z)  =  0.2315
\end{eqnarray*}

\ni and $n_f$ is the number of quark/lepton families.

The inverse couplings are plotted in Fig.1a as a function of
$\sqrt s$.  As observed by Georgi and Glashow [2] all three
couplings seem to reach the same value at an energy $\sqrt s
\simeq 10^{14}$ GeV which is referred to as the \lq\lq Grand
Unification Scale". If the couplings evolve according to the
minimal supersymmetric model (MSSM) much better agreement is
obtained, and within present uncertainties, the constants meet
exactly at $\sqrt s = 10^{15}$ GeV, as shown in Fig.1b. [3].

The gravitational constant depends on the interaction energy as
well. Consider two protons moving against each other in the
laboratory frame with velocity $\beta$ and energy $\gamma m_p$.
The gravitational coupling in this case takes the form

\begin{equation}
 \alpha_G(\sqrt s)  =
\frac{G_Nm_p^2}{\hbar c} (2\gamma^2-1)
\end{equation}

\ni The factor of $2\gamma^2$ arises because both energy $\gamma
m_p$ and momentum $\gamma\beta m_p$ couple; see for instance [4].
The c.m. collision energy is $s=4\gamma^2m_p^2$, so Eq.(1) can be
written as

\begin{equation}
\alpha_G(\sqrt s) = \frac{G_Nm_p^2}{\hbar c}\left[\frac{s}{2m_p^2}
-1\right]
\end{equation}

\ni valid for $s \geq 4m_p^2$. Numerically

\begin{equation}
 \frac{G_Nm_p^2}{\hbar c} =  \left(\frac{m_p}{M_P}\right)^2 = 0.59
 \times 10^{-38}
 \end{equation}

\ni where $M_p$ is the Planck Mass.  When $\sqrt{s/2} = M_p$, then
$\alpha_G$ becomes unity.

If all four forces can be derived from a single gauge group then
the gravitational coupling should reach the common value

\begin{equation}
1/\alpha_G = 1/\alpha_1 = 1/\alpha_2 = 1/\alpha_3 \simeq 42
\end{equation}

\ni at the unification scale $\sqrt s = 10^{15}$ GeV. [5].  We can
achieve this by modifying the classical evolution of $\alpha_G(s)$
by a logarithmic term, of the form

\begin{equation}
\alpha_G (\sqrt s) = \left(\frac{m_p}{M_P}\right)^2
\left[\left(\frac{s}{2m_p^2}\right)^{1+b\ell n(s/4m^2)} -1\right]
\end{equation}

\ni The coefficient $b$ is found, by the requirement of Eq.(4), to
have the value

\begin{equation}
b = 0.00340
\end{equation}

We can now obtain the gravitational coupling at any given
interaction energy.  There are two possibilities for testing this
hypothesis:

In the first case one tries to measure the gravitational effect at
LHC energies.  Here there is a significant difference in the
couplings

\begin{eqnarray*}
\alpha_{GC} (\sqrt s = 1.4 \times 10^4\ {\rm GeV}) & = & 6.6
\times 10^{-31} \qquad\qquad {\rm classical}\\
\\
\alpha_{GR}(\sqrt s = 1.4 \times 10^4\ {\rm GeV}) & = & 2.0 \times
10^{-30} \qquad\qquad {\rm running}
\end{eqnarray*}

\ni However the gravitational effects are extremely small and are
dominated by the much larger electromagnetic force [6].  New
detector technology [7] may make it worthwhile to re-examine such
experiments.

The other possibility is to test the deviation from the classical
behavior at low energies but with macroscopic bodies.  The
classical correction in this case leads to

\begin{equation}
\alpha_{GC}(\beta) = \left(\frac{m_p}{M_P}\right)^2 (1+2\beta^2)
\end{equation}

\ni while for the model of Eq.(5)

\begin{eqnarray}
\alpha_{GR}(\beta) &\simeq & \left(\frac{m_p}{M_P}\right)^2
\left\{\left[2(1+\beta^2)\frac{}{}\right]^{1+b\beta^2}-1\frac{}{}\right\}\nonumber \\
\\
& \simeq & \alpha_{GC}(\beta) [1+x]\nonumber
\end{eqnarray}

\ni with

\begin{equation}
x = 2^{1+b\beta^2} -2 \sim \sqrt 2 b\beta^2
\end{equation}

For a satellite in earth orbit

$$\beta^2 = 0.7 \times 10^{-9}$$

\ni whereas for a close solar orbit

$$\beta^2 = 0.2 \times 10^{-5}$$

\ni In this latter case, the effect of the running coupling
constant is

$$x \simeq 10^{-8}$$

\ni which could be measurable as a difference in the predicted
orbital dynamics, beyond the effects of classical general
relativity [8].

\vspace{2in}

 \ni {\bf References}

\begin{enumerate}

\item M.E. Peskin and D.V. Schroeder \lq\lq An Introduction to
Quantum Field Theory", Addison-Wesley (1995).

\item H. Georgi and S.L. Glashow, Phys. Rev. Lett. \underline{32},
438 (1974).

\item F. Wilczek in \lq\lq Critical Problems in Physics",
Princeton University Press (1997).

\item A.C. Melissinos, Il Nuovo Cimento, \underline{62B}, 190
(1981).

\item I heard of this suggestion in a talk by F. Wilczek at the
symposium in honor of N.P. Samios held at BNL, May 2002.

\item P. Reiner et al., Physics Letters \underline{B176}, 233
(1986).

\item Ph. Bernard et al., Review of Scientific Instruments
\underline{72}, 2428 (2001).

\item See for instance T.P. Krisher et al., Phys. Rev. Lett.
\underline{64}, 1322 (1990).

\end{enumerate}

\end{document}